\documentstyle[aps,prl]{revtex}
\title{\bf An efficient localized basis set for quantum Monte Carlo
calculations on condensed matter } 
\author{D. Alf\`{e}$^{1,2}$ and
M. J. Gillan$^2$
\smallskip \\
$^1$Department of Earth Sciences, University College London \\ Gower
Street, London WC1E~6BT, UK
\smallskip \\
$^2$Department of Physics and Astronomy, University College London \\
Gower Street, London WC1E~6BT, UK}

\begin{document}

\maketitle

\begin{abstract}
We present an efficient scheme for representing many-body
wavefunctions in quantum Monte Carlo (QMC) calculations. The scheme is
based on B-splines (blip functions), which consist of localized cubic
splines centred on the points of a regular grid. We show that
blip functions are unbiased, systematically improvable, and
conveniently obtained from any standard plane-waves density functional
theory (PW-DFT) code, and therefore provide a convenient and natural
interface between PW-DFT and QMC calculations. We present tests on a
16-atom system of Si in the $\beta$-tin structure, and on 2- and 8-
atoms systems of MgO in the NaCl structure. We show that already with
such small systems the speed-up of blip functions with respect to
plane-waves is between one and two order of magnitudes, without
compromising the accuracy.
\end{abstract}

The quantum Monte Carlo (QMC) technique~\cite{foulkes01} is becoming
ever more important in the study of condensed matter, with recent
applications including the reconstruction of semiconductor
surfaces~\cite{healy01}, the energetics of point defects in
insulators~\cite{hood03}, optical excitations in
nanostructures~\cite{williamson02}, and the energetics of organic
molecules~\cite{aspuru04}.  Although its demands on computer power are
much greater than those of widely used techniques such as density
functional theory (DFT), its accuracy is also much greater for most
systems. With QMC now being applied to large complex systems
containing hundreds of atoms, a major issue is the basis set used to
represent the many-body wavefunctions. We propose and test here a
basis set that has many of the properties of plane waves, currently
widely used in density functional theory (DFT) calculations, but with
the advantage of being localized.

In QMC, the trial many-body wavefunction $\Psi_{\rm T} ( {\bf r}_1 ,
\ldots {\bf r}_N )$ consists of a Slater determinant $D$---or more
generally a linear combination of Slater determinants---of
single-electron orbitals $\psi_n ( {\bf r}_i )$ multiplied by a
parameterized Jastrow correlation factor $J ( {\bf r}_1 , \ldots {\bf
r}_N )$. In variational Monte Carlo (VMC), $J$ is `optimized' by
varying its parameters so as to reduce the variance of the `local
energy' $\Psi_{\rm T}^{-1} \left( \hat{H} \Psi_{\rm T} \right)$, where
$\hat{H}$ is the many-electron Hamiltonian. Since VMC by itself is not
usually accurate enough, the optimized $\Psi_{\rm T}$ produced by VMC
is used in diffusion Monte Carlo (DMC), which achieves the exact
ground state within the fixed nodal structure imposed by the Slater
determinant $D$. At each QMC step it is necessary to evaluate
$\Psi_{\rm T} ( {\bf r}_1 ,\ldots {\bf r}_N )$ in each of the replicas
(QMC ``walkers''), which involves the evaluation of the single
electron orbitals $\psi_n ( {\bf r}_i )$. A crucial issue in the
efficiency of the calculations is therefore the representation of
$\psi_n ( {\bf r}_i )$. One common approach is using plane waves
(PW). The big advantages of PW are that their accuracy is
systematically improvable (by increasing the PW cutoff) and they are
unbiased. Moreover, many DFT codes are written in terms of PW, so the
technology is highly developed and easily accessible. However, PW are
not well suited for QMC calculations because for the evaluation of
each $\psi_n ( {\bf r}_i )$ one has to sum over all PW in the
system. Since this has to be done for $M$ orbitals and $N$ electrons,
with $M$ proportional to $N$, the cost of evaluating the many-body
wavefunction involves $O(N^3)$ operations, with a prefactor depending
on PW cutoff, which can be very large for 'hard' systems, like
MgO. The storage required for a PW representation is proportional to
$N^2$.

This problem with PW can be overcome by using localized basis
sets. One possibility is to use gaussians, but the drawback is that
they are biased and generally difficult to improve systematically. An
option which combines the best of both worlds, is to use a B-spline
basis (blip functions), already proposed for $O(N)$ DFT
calculations~\cite{hernandez97}. Here we propose and test the use of
blip functions in QMC calculations. We will show that blip functions
share all the advantages of PW, i.e. are systematically improvable and
unbiased. They are also closely connected with PW, and can therefore
be obtained from PW-DFT codes. However, they are localized, therefore
the evaluation of each orbital $\psi_n ( {\bf r}_i )$ has a cost which
is independent of the size of the system and indeed of blip-grid
spacing (connected to the PW cutoff). The storage required for blip
functions is not much worse than PW and has the same $O(N^2)$ scaling.

As described in detail elsewhere~\cite{hernandez97}, the blip
functions consist of localized cubic splines centred on the points of
a regular grid, each function being non-zero only inside a region
extending two grid spacings in each direction from its centre. For a
cubic grid spacing $a$, the blip function $\Theta_s({\bf r})$ centred
on the grid point at position ${\bf R}_s = (X_s,Y_s,Z_s)$ is given by:
\begin{equation}
\Theta_s({\bf r}) = \phi((x - X_s)/a)\phi((y - Y_s)/a)\phi((z- Z_i)/a),
\end{equation}
where $\phi(\xi)$ is:
\begin{eqnarray}
\phi(\xi)   = 1 - \frac{3}{2}\xi^2 + \frac{3}{4}|\xi|^3 \phantom{pollo} 0 \le |\xi| \le 1 \nonumber \\
\phantom{\phi(\xi)} = \frac{1}{4}(2 - |\xi|)^3 \phantom{polloxxxx} 1 \le |\xi| \le 2 \nonumber \\
\phantom{\phi(\xi)} = 0 \phantom{polloxxxxxxxxxxxxxx} |\xi| \ge 2.
\end{eqnarray}
The function and its first two derivatives are continuous,
discontinuities appear only in the third derivative, and all higher
derivatives are zero. Each single particle orbital is then represented as:
\begin{equation}\label{eqn:blip}
\psi_n({\bf r}) = \sum_s a_{ns} \Theta_s({\bf r}).
\end{equation}
For any position ${\bf r}$, there are only 64 non-zero blip functions,
whatever the nature and size of the system, so that the number of
operations to compute $\psi_n({\bf r})$ is the same for any material.

The close relationship between B-splines and PW has been discussed
elsewhere~\cite{hernandez97}. In the PW representation, the single
particle orbitals are given by:
\begin{equation}
\psi_n({\bf r}) = \sum_{\bf k} c_{n{\bf k}} e^{i {\bf k} \cdot {\bf r}} 
\end{equation}
where the wavevectors ${\bf k}$ go over the reciprocal lattice vectors
of the superlattice, with $k$ less than the PW cutoff
$k_{max}$. The relationship between the PW coefficients $c_{n{\bf k}}$
and the blip coefficients $a_{n{\bf s}}$ can be understood by
considering blip waves $\chi_{\bf k}(\bf r)$ defined by:
\begin{equation}
\chi_{\bf k}({\bf r}) = \sum_s e^{i{\bf k \cdot R}_s}\Theta_s({\bf r}).
\end{equation}
For small ${\bf k}$, the $\chi_{\bf k}(\bf r)$ are essentially
identical to plane waves $\exp({\bf k} \cdot {\bf r})$, apart from a
${\bf k}$-dependent factor $\gamma_{\bf k}$:
\begin{equation}
e^{i{\bf k \cdot r}} \simeq \gamma_{\bf k} \chi_{\bf k}({\bf r}).
\end{equation}
The factor $\gamma_{\bf k}$ is the Fourier transform of a single blip
function $\Theta(\bf r)$ and is given by $\gamma_{\bf k} =
\gamma_{k_x} \gamma_{k_y} \gamma_{k_z}$, where ${\bf k} = (k_x,k_y,k_z)$ and 
\begin{equation}
\gamma_k = \frac{3}{k^4}(3 - 4 \cos k + \cos 2k).
\end{equation}
At larger ${\bf k}$, the $\chi_{\bf k}(\bf r)$ differ significantly
from $\exp({\bf k} \cdot {\bf r})$, as they must, because $\chi_{\bf
k}(\bf r)$ is periodic in ${\bf k}$ space: the number of independent
$\chi_{\bf k}(\bf r)$ functions is equal to the number of sites on the
blip grid.

There is a ``natural'' choice of blip grid spacing $a$, given by
$a=\pi/k_{\rm max}$. With this choice, the region $k \approx k_{max}$
where blip-waves and plane-waves differ most is the region where the
plane waves coefficients $c_{n{\bf k}}$ are very small. However, the
precision with which blip-waves reproduce plane-waves in the region
$k<k_{max}$ can always be improved by refining the blip grid.

The procedure to obtain the blip coefficients $a_{ns}$ from the
plane-wave coefficients of orbitals $\psi_n({\bf r})$ obtained from a
DFT calculation is straightforward. For the relationship between
blip-waves and plane-waves (see eqn. 3, 4 and 7), it follows that
\begin{equation}
a_{ns} = \sum_{\bf k} c_{n{\bf k}} \gamma_{\bf k} e^{i {\bf k} \cdot {\bf R}_s}.
\end{equation}
The coefficients $a_{ns}$ can therefore be evaluated using fast
Fourier transform routines.

We have implemented blip functions in the appropriately modified {\sc
casino} code~\cite{needs04}. To test the implementation we present now
three cases in which we compare the energy and the standard deviation
in VMC and DMC calculations performed using PW or blip-functions
representations of the single-particle orbitals.  Calculations with
blip functions are presented for two values of the grid spacing, the
{\em natural grid} spacing $a=\pi/k_{\rm max}$ and a two times finer
grid obtained with $a=\pi/2k_{\rm max}$. Results are reported in
Table~\ref{tab:blip_tests}. All calculations have been performed at
the $\Gamma$ point.

The first case is a 16-atom cell of silicon in the $\beta$-tin
structure. The single-particle orbitals have been obtained using the
{\sc pwscf} code~\cite{PWSCF}, with Hartree-Fock pseudo-potentials
(p-channel chosen to be the local part) and PW cutoff energy of 15 Ry.
VMC calculations are reported for $3.2\times 10^6$ steps of length 1
a.u. in all three cases. No Jastrow factor has been used for these VMC
calculations. DMC calculations have been performed using 320 walkers
for 10100, 12700 and 10100 steps of length 0.03 a.u. for PW and blip
functions calculations with the coarse and the fine grid spacing
respectively. Diffusion to the ground state is already achieved after
$\approx 100$~steps.  In VMC the natural grid is not dense enough for
this system, with the largest difference being in the kinetic energy
(of $\approx 0.06$ eV/atom). The standard deviation on the energy is
also slightly larger. However, with the fine grid the blip functions
results agree identically with the PW ones within a statistical error
of only a few meV/atom. In Table~\ref{tab:blip_tests} we also report
the time taken to perform one VMC step on an Origin 3000
machine. Already for this small system, with such a modest PW cutoff,
the speed-up with blip functions is almost a factor of 6. The timings
between the two blip-function calculations should in principle be
identical, the small difference between the two is probably due to the
larger sparsity in memory of the blip coefficients $a_{n{\bf l}}$ for
the case with a finer grid, and we found that this is machine
dependent. For DMC the computational speed-up is more that a factor of
10, and the energy is already correct with the natural grid, which
means that the nodal surface is essentially the same as the
PW one already with the natural grid.

The second test we performed was a perfect crystal of MgO in its zero
pressure NaCl structure. The unit cell in this case contained only 2
atoms and had face-centred-cubic (f.c.c.) geometry. Single-particle
orbitals were obtained again using the {\sc pwscf} code, with
Hartree-Fock pseudo-potentials (d-channel chosen as the local part for
both Mg and O) and a PW cutoff of 200 Ry. No Jastrow factor has been
used in these calculations. VMC calculations have been done with
$3.36\times 10^7$, $1.6\times 10^8$ and $1.6\times 10^8$ steps of
length 0.3 a.u. for PW and the two blip functions cases. DMC
calculations have been performed using 1600 walkers for
$10.79\times10^{4}$, $23.98\times10^{4}$, $11.53\times10^{4}$ steps of
length 0.005 a.u. for the three cases~\cite{time_step}. Diffusion to
the ground state is achieved after the first few hundred
steps. Similarly to the previous case, blip functions VMC energies and
standard deviation agree identically with those obtained using PW for
the dense blip grid, and the speed-up obtained with blip functions is
more than a factor of 10. DMC energies are also in this case correct
already when the coarse grid is used, but the variance is
significantly improved when the fine grid is used.

Finally, the third test consists of the same MgO crystal simulated in
a simple cubic (s.c.) cell containing 8 atoms. Single-particle
orbitals were obtained in analogy to the previous case, i.e. same pseudo-potential
and same PW cutoff of 200 Ry. No Jastrow factor has been used. VMC
calculations have been done with $0.32\times 10^6$, $1.6\times 10^6$
and $1.6\times 10^6$ steps of length 0.3 a.u. for PW and the two blip
functions cases. The important thing to notice in this case is the
speed-up obtained with blip functions, which is over two order of
magnitudes. We have not attempted DMC calculations as they would be
impractical for the PW case. 

We note that despite we have chosen to use PW cutoffs of 15 Ry and 200
Ry for Si and MgO respectively, we found that by using much larger
cutoff energy (typically 32 Ry for Si and 500 Ry for MgO) the variance
of the energy can be further significantly improved. Of course,
increasing the PW cutoff leads to a direct increase in the PW
computational time, but has hardly any effect in the calculations
which employ blip functions. We have also found that by using much
larger PW cutoff the blip functions natural grid is already accurate
enough, as expected.

We have presented here a robust and efficient scheme based on
B-splines to represent the trial wavefuntions in QMC calculations. We
have shown that this scheme shares all the advantages of plane-waves,
but offers a much better scaling behaviour with respect to the number
of atoms in the system and the hardness of the pseudo-potentials used
in the calculations. This scheme has been implemented in the {\sc
casino}~\cite{needs04} code, and we have presented tests on three
different cases. The largest system considered here (in terms of
number of plane waves) was an MgO crystal in the NaCl structure
simulated with a s.c. unit cell containing 8 atoms. We have shown that
already for this relatively small system the speed-up obtained using
blip functions is over a factor of 100. Since B-splines can easily be
obtained from PW, they also provide a natural and convenient interface
between QMC and PW-DFT codes. Moreover, this technique can be used in
conjuction with ``linear-scaling'' techniques for QMC calculations, as
reported elsewhere~\cite{williamson01,alfe04}. We conclude by noting
that we are now attempting to calculate the formation energy of a
Schottky defect in MgO using a cell containing 54 atoms. This
calculation would be impossible to perform if we had to use PW
(results will be reported elsewhere~\cite{alfe04a}).

\smallskip
DA acknowledges support from the Royal Society, and also thanks the
Leverhulme Trust and the CNR for support. The authors are indebted to
R.J.~Needs, M.D.~Towler and N.D.~Drummond for advice and technical
support in the use of the {\sc casino} code, and for providing us with
the pseudopotentials used in this work. Allocation of computer time ar
the CSAR service were provided by the Mineral Physics Consortium
(NERC grant GST/02/1002).

\begin{table}
\begin{tabular}{lcccc}
\hline
  &       &   PW    &  Blips($a=\pi/k_{\rm max}$)   &   Blips($a=\pi/2k_{\rm max}$) \\ 
\hline
\multicolumn{2}{l}{Si $\beta$-tin, 16 atoms} & & & \\
VMC & & & & \\
& $E_{\rm kin}$ & 43.864(3) & 43.924(3) & 43.862(3) \\
& $E_{\rm loc}$ & 15.057(3) & 15.063(3) & 15.058(3) \\
& $E_{\rm nl}$  & 1.533(3)  & 1.525(3)  & 1.535(3) \\
& $E_{\rm tot}$  & -101.335(3) 4.50 & -101.277(3) 4.74 & -101.341(3) 4.55 \\
& T (s/step) & 1.83 & 0.32 & 0.34 \\
DMC & & & & \\
& $E_{\rm tot}$  & -105.713(3) 2.29 & -105.711(4) 2.95 & -105.715(4) 2.38 \\
& T (s/step) & 2.28 & 0.21 & 0.25 \\
\hline
\multicolumn{2}{l}{MgO-NaCl, 2 atoms, f.c.c. cell} & & & \\
VMC & & & & \\
& $E_{\rm kin}$ & 199.449(24)  & 199.465(15) & 199.418(15) \\
& $E_{\rm loc}$ & -239.899(27) & -239.861(15) & -239.855(15)\\
& $E_{\rm nl}$  & -26.906(12)  & -26.889(8) & -26.902(8) \\
& $E_{\rm tot}$  & -224.527(4) 28.7 &  -224.465(3) 35.8 & -224.523(2) 28.3 \\
& T (s/step) & $101 \times 10^{-3}$ &  $8.3 \times 10^{-3}$ & $8.9 \times 10^{-3}$ \\
DMC & & & & \\
& $E_{\rm tot}$  & -228.429(10) 22.1 & -228.433(7) 28.9 & -228.427(9) 22.3 \\
& T (s/step) & $89 \times 10^{-3}$ & $7.1 \times 10^{-3}$ & $7.5 \times 10^{-3}$  \\
\hline
\multicolumn{2}{l}{MgO-NaCl, 8 atoms, s.c. cell} & & & \\
VMC & & & & \\
& $E_{\rm kin}$ & 178.349(49)  & 178.360(22) & 178.369(22) \\
& $E_{\rm loc}$ & -225.191(50) & -225.128(24) & -225.177(23)\\
& $E_{\rm nl}$  &  -17.955(25) & -17.974(11) & -17.976(11) \\
& $E_{\rm tot}$  & -227.677(8) 14 &  -227.648(4) 15 & -227.669(4) 14.5 \\
& T (s/step) & 7.8 &  $5.6 \times 10^{-2}$ & $7.1 \times 10^{-2}$ \\
\hline
\end{tabular}
\caption{Comparisons of the various components of the total energy (in
eV/atom) and timings between VMC and DMC for a 16 atoms Si system in
the $\beta$-tin structure, and an MgO crystal in the NaCl
structure. Standard deviation of the total energy (eV/atom) is also
reported beside the total energy.  The MgO crystal has been simulated
using a 2 atoms face-centred-cubic cell and an 8 atoms simple cubic
cell.  PW calculations have been performed with a cutoff energy of 15
Ry for Si and 200 Ry for MgO. Blips calculations have been performed
using two different grid spacing: $a=\pi/k_{\rm max}$ and
$a=\pi/2k_{\rm max}$, where $k_{\rm max}$ is the modulus of the
largest PW vector.}\label{tab:blip_tests}
\end{table}


\begin{thebibliography}{99}
\bibitem{time_step} With the choice of 1 a.u. and 0.3 a.u. for VMC
time steps for Si and MgO respectively the probability of accepting
the move is roughly 50 \% in both cases. The requirement of a shorter
time step for MgO is due to the higher spacial variability of the
wavefunctions for this system, also reflected by the need of a higher
PW cutoff. For DMC, the time steps of 0.03 a.u. and 0.005 a.u. for Si
and MgO respectively results in an undetectable time step error within
the statistical error of a few meV/atom.
\end{thebibliography}
\end{document}